\documentclass{article}

    \PassOptionsToPackage{numbers, compress}{natbib}

\usepackage{natbib}

    \usepackage[preprint]{neurips_2023}

\usepackage[disable=footnote]{microtype}
\usepackage[utf8]{inputenc} 
\usepackage[T1]{fontenc}    
\usepackage{hyperref}       
\usepackage{url}            
\usepackage{booktabs}       
\usepackage{amsfonts}       
\usepackage{nicefrac}       
\usepackage{xcolor}         
\usepackage{graphicx}
\usepackage{wrapfig}
\usepackage{tabularx, booktabs}
\usepackage{multirow}
\usepackage{caption}
\usepackage{amsmath}
\usepackage{setspace}

\title{Synthetic Tumor Manipulation: \\
With Radiomics Features}

\author{%
  Inye Na, Jonghun Kim, Hyunjin Park\thanks{Corresponding author} \\ \\
  Department of Electrical and Computer Engineering, Sungkyunkwan University, Korea\\
  \texttt{\{niy0404, iproj2, hyunjinp\}@skku.edu} \\
}

\begin{document}

\maketitle

\begin{abstract}

We introduce RadiomicsFill, a synthetic tumor generator conditioned on radiomics features, enabling detailed control and individual manipulation of tumor subregions. This conditioning leverages conventional high-dimensional features of the tumor (i.e., radiomics features) and thus is biologically well-grounded. Our model combines generative adversarial networks, radiomics-feature conditioning, and multi-task learning. Through experiments with glioma patients, RadiomicsFill demonstrated its capability to generate diverse, realistic tumors and its fine-tuning ability for specific radiomics features like `Pixel Surface' and `Shape Sphericity'. The ability of RadiomicsFill to generate an unlimited number of realistic synthetic tumors offers notable prospects for both advancing medical imaging research and potential clinical applications.

\end{abstract}

\section{Introduction}

Medical imaging faces challenges in data collection due to concerns related to patient privacy and high labeling costs. However, effective neural network training necessitates a large and diverse dataset \cite{russakovsky2015imagenet}. Recent research trends aim to overcome this challenge through the generation of synthetic data using generative models such as generative adversarial networks (GANs) and diffusion \cite{hernandez2022synthetic, akbar2023brain, kebaili2023deep}.

Liu et al. \cite{liu2023partial} proposed a partial convolution GAN that synthesizes lesions using predefined lesion contours and textures. Kidder \cite{kidder2023advanced} demonstrated the generation of magnetic resonance imaging (MRI) for low-grade glioma using the text-to-image model, Dreambooth, emphasizing the preservation of patient anonymity. Furthermore, Shin et al. \cite{shin2018medical} introduced a method that employs GANs to generate synthetic abnormal MRIs by merging a segmented brain mask from a normal brain MRI with a predefined tumor label.

However, these existing studies present limitations. These methods tend to generate tumors with a uniform intensity distribution and do not allow users to change specific features of the synthetic tumors. Notably, no method provides individual control of features within the tumor region. 

In response, we introduce RadiomicsFill, a model that generates tumors using radiomics features, high-dimensional conventional features to describe the tumor \cite{aerts2014decoding},  as conditions. 
The advantages of RadiomicsFill include;   \textbf{
1) Detail \& Realism:} 
With high-dimensional radiomics features based on shape, histogram, and texture as conditions, it enables the fine-tuning of tumors with enhanced detail and realism. \textbf{
2) Subregion Control:}
The model facilitates the individualized manipulation of radiomics features within tumor regions, allowing users to generate sharp and diverse synthetic tumors.  
\textbf{
3) Anatomical Consideration: }
The model considers the anatomical structures surrounding the tumor during inpainting. 
\textbf{
4) BackgroundFill: }
Removes existing tumors and creates them in new locations.

\begin{figure}
  \centering
  \includegraphics[width=1.0\linewidth]{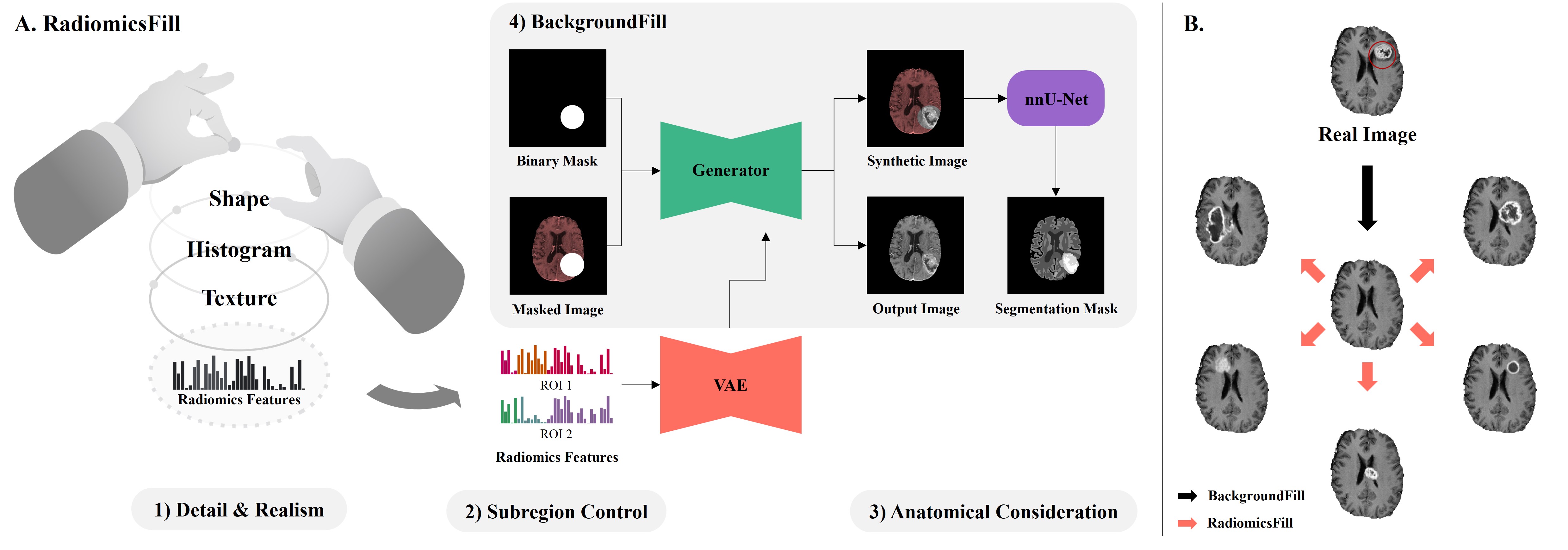}
  \caption{\textbf{A.} Overview of RadiomicsFill.  \textbf{B.} Illustration of a tumor removed from real data using BackgroundFill and the generation of new tumors at various locations with RadiomicsFill.}
  \label{fig1}
\end{figure}

With RadiomicsFill, researchers can produce diverse realistic tumors, as illustrated in Figure \ref{fig1}B. Our model, when used to specifically generate underrepresented tumors, can be a potential solution to tackle data imbalances. Its utility could extend to areas like pre-treatment planning and monitoring post-treatment tumor changes.

\section{Methods}
\label{gen_inst}

\subsection{Dataset}

For our experiments and evaluations, we utilized the Brain Tumor Segmentation (BraTS) 2021 training data \cite{brats2021_1, brats2021_2, brats2021_3}, which offers glioma datasets with three types of tumor masks (i.e., necrotic, enhancing tumor, and edema). We used contrast T1-weighted (cT1) MRI, segmentation labels. We partitioned the data of 1,251 patients: 1,000 for training, 76 for validation, and 100 for testing. Data from 75 patients were excluded due to loading and radiomics feature extraction errors.

\subsection{RadiomicsFill}

The overall structure of RadiomicsFill is depicted in Figure \ref{fig1}A. Our model combines GAN, radiomics-feature conditioning, and multi-task learning. The backbone network for both the Generator and Discriminator was chosen from DeepFillv2 \cite{yu2019free}. The model takes binary masks, MRI images, and radiomics features as inputs. The mask is a circular mask whose diameter is the Maximum Diameter of the radiomics feature that reflects the size of the tumor. For tumor inpainting, a cross-attention module was incorporated to condition on the radiomics features. The latent vector z, derived after passing the radiomics feature through an encoder of the variational autoencoder (VAE), acts as this condition. Consequently, even with identical radiomics features, the model can generate various tumors. Radiomics features consist of 67 attributes for each tumor mask: 9 shape-based, 18 histogram-based, and 40 texture features, extracted using PyRadiomics \cite{pyradiomics}.

To ensure RadiomicsFill reflects anatomical considerations, we incorporated a pretrained nnU-Net \cite{nnunet} to predict brain-related segmentation labels for cerebrospinal fluid (CSF), white matter, gray matter, and tumor in a multi-task learning setup. Labels for CSF, white matter, and gray matter were derived from cT1 MRI  using FastSurfer \cite{henschel2020fastsurfer}. Consequently, tumors generated by RadiomicsFill consider adjacent regions for enhanced realism and adhere closely to anatomical details.

RadiomicsFill's entire network is optimized through a combination of adversarial loss, VAE loss, non-mask loss, and mask loss. The non-masked region, which should be reconstructed identically to the input, is assessed using the L1 loss. In contrast, the mask loss, which consists of perceptual and style losses based on the middle layers of the nnU-Net, allows the network to capture mid-level features and style details specific to the masked regions.

\subsection{BackgroundFill}

Our BackgroundFill model, illustrated in Figure \ref{fig1}A (4), aimed at removing tumors from MRI images, follows a structure similar to RadiomicsFill, excluding the VAE. It is trained using tumor-free slices from the RadiomicsFill training set. For the training, circular masks with random diameters were used.

\begin{table}
  \caption{Radiomics feature similarity between synthetic and real data.}
  \label{tab1}
  \centering
  \resizebox{\textwidth}{!}{%
  \begin{tabular}{ccccccccc}
    \toprule
    & \multicolumn{4}{c}{ROI1} & \multicolumn{4}{c}{ROI2} \\
    \cmidrule(r){2-5} \cmidrule(r){6-9}
     & Shape & Histogram & GLCM & GLSZM & Shape & Histogram & GLCM & GLSZM \\
    \midrule
    Cosine Similarity & 0.8548& 0.6592& 0.7301& 0.9137& 0.9522& 0.9722& 0.8337& 0.9277\\
    Pearson Correlation & 0.8295*& 0.6369*& 0.7220*& 0.8942*& 0.9404*& 0.9702*& 0.8279*& 0.9101*\\
    Spearman Correlation & 0.9439*& 0.9142*& 0.9065*& 0.9834*& 0.9900*& 0.9637*& 0.9522*& 0.9882*\\
    \bottomrule
  \end{tabular}
  }

  \footnotesize 
  \parbox{\textwidth}{%
    \textit{ROI1 represents the necrotic region. ROI2 typically represents the enhancing tumor, but for the shape feature, it combines both the necrotic region and enhancing tumor. ROI; region of interest, GLCM; gray level co-occurrence matrix texture features, GLSZM; gray level size zone matrix texture features; * indicates p-value < 0.0001}}
\end{table}

\section{Experimental Results}

\begin{wrapfigure}{r}{0.5\linewidth}
  \centering
  \vspace{-1ex}
  \includegraphics[width=1.0\linewidth]{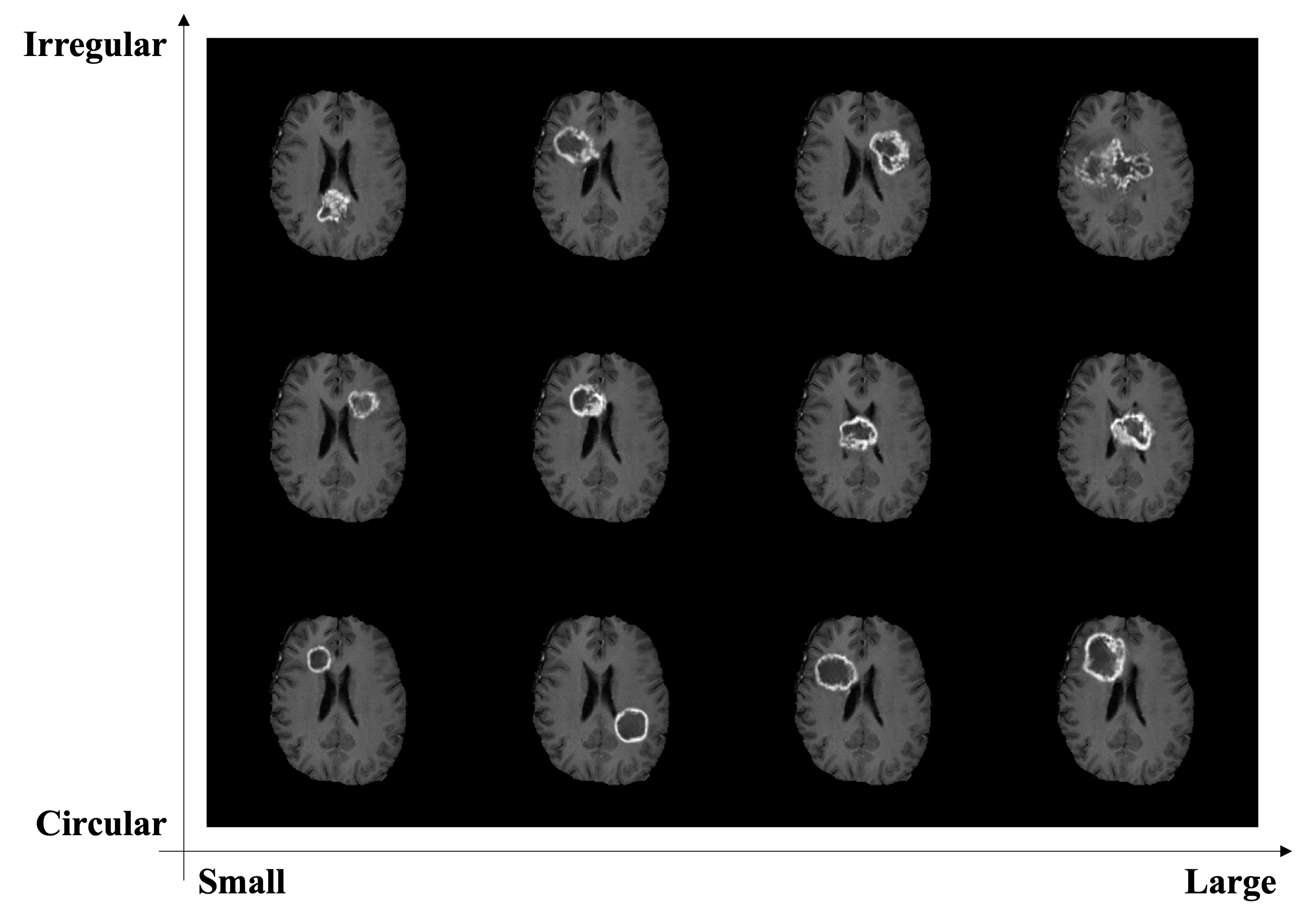}
  \caption{\textbf Visualization of synthetic tumors for the same input image varying the `Pixel Surface' and `Shape Sphericity' radiomics features of the combined necrotic and enhancing tumor region , displayed in a 2D grid.}
  \label{fig2}
  \vspace{-1ex}
\end{wrapfigure}

To evaluate RadiomicsFill's effectiveness, we assessed the radiomics feature similarity and conducted a qualitative analysis of the generated synthetic tumors. The feature matching, detailed in Table \ref{tab1}, shows a high degree of similarity between the synthetic and real data, indicating the model's ability to accurately reflect input features. For this, the segmentation mask output from nnU-Net was employed for feature extraction from synthetic data.

Figure \ref{fig2} visualizes synthetic tumors, each generated from a single input image but conditioned on two different radiomics features, plotted against the `Pixel Surface' and `Shape Sphericity' of the combined necrotic and enhancing tumor region as axes. This visualization facilitates a qualitative assessment of RadiomicsFill's capacity to fine-tune the generation of diverse tumors, ranging from small to large and circular to irregular.

\section{Conclusion}

In this study, we present RadiomicsFill, a model adept at generating synthetic tumors, conditioned on radiomics features. This innovation allows for an intricate manipulation of individual tumors. Validation using glioma patients' cT1 MRIs confirmed the model’s capability to produce realistic synthetic tumors, aligning with specified input conditions and being effectively identified by segmentation networks. Additionally, experiments involving the generation and comparison of synthetic tumors from the same input image, with subtle variations in `Pixel Surface' and `Shape Sphericity', demonstrated RadiomicsFill’s capacity for fine-tuning to create realistic tumors. This underscores the potential of RadiomicsFill in planning and monitoring responses to treatment, showcasing its applicability in real-time clinical scenarios. We anticipate that the extended application of this model will significantly benefit medical imaging research and offer valuable insights for clinical applications.

\section*{Acknowledgment}

This study was supported by National Research Foundation (NRF-2020M3E5D2A01084892), Institute for Basic Science (IBS-R015-D1), ITRC support program (IITP-2023-2018-0-01798), AI Graduate School Support Program (2019-0-00421), ICT Creative Consilience program (IITP-2023-2020-0-01821), and the Artificial Intelligence Innovation Hub program (2021-0-02068).

\section*{Potential Negative Societal Impacts}

In our assessment, the methodology in this paper offers no negative societal impacts. We used anonymized data respecting  patient privacy. RadiomicsFill primarily aims to advance medical imaging studies but can also support clinicians in tasks like pre-treatment planning and post-treatment monitoring. It's crucial to note that its purpose is to assist, not replace, physician decisions, minimizing potential misuse. Overall, our approach stands as both innovative and ethically responsible.

\small
\bibliographystyle{unsrt}
\bibliography{ref}

\begin{thebibliography}{10}

\bibitem{russakovsky2015imagenet}
Olga Russakovsky, Jia Deng, Hao Su, Jonathan Krause, Sanjeev Satheesh, Sean Ma, Zhiheng Huang, Andrej Karpathy, Aditya Khosla, Michael Bernstein, et~al.
\newblock Imagenet large scale visual recognition challenge.
\newblock {\em International journal of computer vision}, 115:211--252, 2015.

\bibitem{hernandez2022synthetic}
Mikel Hernandez, Gorka Epelde, Ane Alberdi, Rodrigo Cilla, and Debbie Rankin.
\newblock Synthetic data generation for tabular health records: A systematic review.
\newblock {\em Neurocomputing}, 493:28--45, 2022.

\bibitem{akbar2023brain}
Muhammad~Usman Akbar, M{\aa}ns Larsson, and Anders Eklund.
\newblock Brain tumor segmentation using synthetic mr images--a comparison of gans and diffusion models.
\newblock {\em arXiv preprint arXiv:2306.02986}, 2023.

\bibitem{kebaili2023deep}
Aghiles Kebaili, J{\'e}r{\^o}me Lapuyade-Lahorgue, and Su~Ruan.
\newblock Deep learning approaches for data augmentation in medical imaging: A review.
\newblock {\em Journal of Imaging}, 9(4):81, 2023.

\bibitem{liu2023partial}
Yingao Liu, Fei Yang, and Yidong Yang.
\newblock A partial convolution generative adversarial network for lesion synthesis and enhanced liver tumor segmentation.
\newblock {\em Journal of Applied Clinical Medical Physics}, 24(4):e13927, 2023.

\bibitem{kidder2023advanced}
Benjamin~L Kidder.
\newblock Advanced image generation for cancer using diffusion models.
\newblock {\em bioRxiv}, pages 2023--08, 2023.

\bibitem{shin2018medical}
Hoo-Chang Shin, Neil~A Tenenholtz, Jameson~K Rogers, Christopher~G Schwarz, Matthew~L Senjem, Jeffrey~L Gunter, Katherine~P Andriole, and Mark Michalski.
\newblock Medical image synthesis for data augmentation and anonymization using generative adversarial networks.
\newblock In {\em Simulation and Synthesis in Medical Imaging: Third International Workshop, SASHIMI 2018, Held in Conjunction with MICCAI 2018, Granada, Spain, September 16, 2018, Proceedings 3}, pages 1--11. Springer, 2018.

\bibitem{aerts2014decoding}
Hugo~JWL Aerts, Emmanuel~Rios Velazquez, Ralph~TH Leijenaar, Chintan Parmar, Patrick Grossmann, Sara Carvalho, Johan Bussink, Ren{\'e} Monshouwer, Benjamin Haibe-Kains, Derek Rietveld, et~al.
\newblock Decoding tumour phenotype by noninvasive imaging using a quantitative radiomics approach.
\newblock {\em Nature communications}, 5(1):4006, 2014.

\bibitem{brats2021_1}
Ujjwal Baid, Satyam Ghodasara, Suyash Mohan, Michel Bilello, Evan Calabrese, Errol Colak, Keyvan Farahani, Jayashree Kalpathy-Cramer, Felipe~C Kitamura, Sarthak Pati, et~al.
\newblock The rsna-asnr-miccai brats 2021 benchmark on brain tumor segmentation and radiogenomic classification.
\newblock {\em arXiv preprint arXiv:2107.02314}, 2021.

\bibitem{brats2021_2}
Bjoern~H Menze, Andras Jakab, Stefan Bauer, Jayashree Kalpathy-Cramer, Keyvan Farahani, Justin Kirby, Yuliya Burren, Nicole Porz, Johannes Slotboom, Roland Wiest, et~al.
\newblock The multimodal brain tumor image segmentation benchmark (brats).
\newblock {\em IEEE transactions on medical imaging}, 34(10):1993--2024, 2014.

\bibitem{brats2021_3}
Spyridon Bakas, Hamed Akbari, Aristeidis Sotiras, Michel Bilello, Martin Rozycki, Justin~S Kirby, John~B Freymann, Keyvan Farahani, and Christos Davatzikos.
\newblock Advancing the cancer genome atlas glioma mri collections with expert segmentation labels and radiomic features.
\newblock {\em Scientific data}, 4(1):1--13, 2017.

\bibitem{yu2019free}
Jiahui Yu, Zhe Lin, Jimei Yang, Xiaohui Shen, Xin Lu, and Thomas~S Huang.
\newblock Free-form image inpainting with gated convolution.
\newblock In {\em Proceedings of the IEEE/CVF international conference on computer vision}, pages 4471--4480, 2019.

\bibitem{pyradiomics}
Joost~JM Van~Griethuysen, Andriy Fedorov, Chintan Parmar, Ahmed Hosny, Nicole Aucoin, Vivek Narayan, Regina~GH Beets-Tan, Jean-Christophe Fillion-Robin, Steve Pieper, and Hugo~JWL Aerts.
\newblock Computational radiomics system to decode the radiographic phenotype.
\newblock {\em Cancer research}, 77(21):e104--e107, 2017.

\bibitem{nnunet}
Fabian Isensee, Paul~F Jaeger, Simon~AA Kohl, Jens Petersen, and Klaus~H Maier-Hein.
\newblock nnu-net: a self-configuring method for deep learning-based biomedical image segmentation.
\newblock {\em Nature methods}, 18(2):203--211, 2021.

\bibitem{henschel2020fastsurfer}
Leonie Henschel, Sailesh Conjeti, Santiago Estrada, Kersten Diers, Bruce Fischl, and Martin Reuter.
\newblock Fastsurfer-a fast and accurate deep learning based neuroimaging pipeline.
\newblock {\em NeuroImage}, 219:117012, 2020.

\end{thebibliography}
\normalsize

\end{document}